\documentclass[print,showpacs,column,11pt,pra]{revtex4}%
\usepackage{amsfonts}
\usepackage{amsmath}
\usepackage{amssymb}
\usepackage{graphicx}%
\begin{document}
\title{Universal quantum computation with electronic qubits in decoherence-free subspace}
\author{X.L. Zhang$^{1,2}$}\thanks{%
xlzhang168@gmail.com; present address: Center for Modern Physics
and Department of Physics, Chongqing University, Chongqing\,\ {\rm
400044}, China{\ }}
\author{M. Feng$^{1}$}
\thanks{mangfeng1968@yahoo.com}
\author{K.L. Gao$^{1}$}
\thanks{klgao@wipm.ac.cn}
\affiliation{$^{1}$State Key Laboratory of Magnetic Resonance and
Atomic and Molecular Physics, Wuhan Institute of Physics and
Mathematics, Chinese Academy of Sciences, Wuhan 430071, China}
\affiliation{$^{2}$Graduate School of the Chinese Academy of
Sciences, Beijing 100049, China}


\begin{abstract}
We investigate how to carry out universal quantum computation
deterministically with free electrons in decoherence-free subspace by using
polarizing beam splitters, charge detectors, and single-spin rotations.
Quantum information in our case is encoded in spin degrees of freedom of the
electron-pairs which construct a decoherence-free subspace. We design building
blocks for two noncommutable single-logic-qubit gates and a logic controlled
phase gate, based on which a universal and scalable quantum information
processing robust to dephasing is available in a deterministic way.

\end{abstract}
\maketitle

\section{Introduction}

Decoherence is one of the main obstacles in building quantum computing
architecture, which yields both operational errors and loss of coherence and
entanglement. To defeat decoherence, people have so far proposed a number of
ideas, for example in \cite{1,2,3,4,5,6,7,8}, where the error avoiding
strategies carried out in decoherence-free subspaces (DFS), compared with
other error correction schemes, are relatively simpler, because they only
require some special encodings immune from certain system-environment
disturbances and no error correction steps are needed.

The present work is focused on a dephasing-free scheme for universal quantum
computation with free electrons. Free electrons have recently been shown to be
available for constructing universal quantum computation \cite{9,10}, in which
the basic idea is the quantum information encoded in electron spins and the
quantum gating assisted by electronic charge detection. Due to independence
between the spin and charge degrees of freedom of the electrons, we may design
a quantum computing architecture with free electrons like the idea used in
linear optical quantum computation \cite{11}. For example, like photons,
electrons could be employed to carry out cluster-state preparation and
multipartite analyzer \cite{12}, and to do entanglement purification
\cite{13}. But actually, there are some intrinsic differences between
electrons and photons. The photons do not interact directly with each other,
but electrons repel due to Coulomb interaction. So the works \cite{9,10} are
based on a screening model in which interaction-free assumption is applicable
to free electrons. Besides, different polarized states are degenerate to
photons, which excludes any possibility of dephasing due to different states
in evolution. While an electronic level would be split in a magnetic field due
to Zeeman interaction. As a result, dephasing is an important detrimental
factor in application of free electrons for quantum computation.

To remove dephasing effects, we used to employ refocusing techniques, which
reverse the time evolution during some elaborately selected periods, and
counteract the influence from dephasing. However, refocusing operation does
not work for magnetic field fluctuation, which happens unpredictably in
experiments. To reduce dephasing, we must work in DFS. \textit{The motivation
of this paper is to design a DFS scheme for free-electron model so that the
free electrons, in addition to above mentioned screening\ assumption, could be
fully modeled as free photons for quantum computation.} We define the encoding
to be $\left\vert 0_{L}\right\rangle =\left\vert 01\right\rangle $ and
$\left\vert 1_{L}\right\rangle =\left\vert 10\right\rangle $, which
constitutes a well-known DFS scheme immune from dephasing induced by the
system-environment interaction in the form of $Z\otimes B$, where
$Z=\sigma_{z}^{1}\oplus\sigma_{z}^{2}$ and B is a random bath operator. We
will assume throughout our scheme that the collective dephasing is dominant
for the electron pairs because the collective errors due to coupling to
environment are generally considered to be the main problem for solid-state
systems at low temperature and dephasing is dominant in the corresponding
class of quantum computing devices. While for other dephasing effects, we may
overcome them with the same DFS encoding by some additional dynamical
decoupling (e.g., 'Bang-Bang') pulses \cite{14,15}. To avoid confusion, we
call $\left\vert 0_{L}\right\rangle $ and $\left\vert 1_{L}\right\rangle $
logic-qubits, and $\left\vert 0\right\rangle $ and $\left\vert 1\right\rangle
$ physical-qubits with 0 (1) representing spin up (down). The four Bell states
in our scheme are thereby expressed as $\left\vert \Psi_{L}^{\pm}\right\rangle
=\left(  \left\vert 0110\right\rangle \pm\left\vert 1001\right\rangle \right)
/\sqrt{2}$ $\equiv\left(  \left\vert 0_{L}1_{L}\right\rangle \pm\left\vert
1_{L}0_{L}\right\rangle \right)  /\sqrt{2}$ and $\left\vert \Phi_{L}^{\pm
}\right\rangle =\left(  \left\vert 0101\right\rangle \pm\left\vert
1010\right\rangle \right)  /\sqrt{2}\equiv\left(  \left\vert 0_{L}%
0_{L}\right\rangle \pm\left\vert 1_{L}1_{L}\right\rangle \right)  /\sqrt{2}$.
Since the logic-qubits are immune from collective dephasing, the dephasing
errors would be considerably suppressed in implementation of our designed
quantum computation.

\section{Building blocks for quantum gates}

We now address in detail how to realize a universal quantum computation in the
DFS by building blocks involving polarizing beam splitters, charge detectors,
and single-spin rotations. The basic components for a universal quantum
computation are two noncommutable single-qubit operations and a nontrivial
two-qubit gate. For an arbitrary logic-qubit state, e.g., $\alpha\left\vert
0_{L}\right\rangle +\beta\left\vert 1_{L}\right\rangle \equiv\alpha\left\vert
01\right\rangle +\beta\left\vert 10\right\rangle $, a single-logic-qubit
rotation $R_{z}$ can be realized by rotating one of the physical-qubits, as
shown in Fig. 1 (a): If we consider a rotation of the first physical-qubit, as
the potential barrier in Fig. 1 (a) makes $\left\vert 1\right\rangle $ be
$e^{i\theta}\left\vert 1\right\rangle $ but $\left\vert 0\right\rangle $
unchanged, we have $\alpha\left\vert 0_{L}\right\rangle +\beta\left\vert
1_{L}\right\rangle \Longrightarrow\alpha\left\vert 01\right\rangle +\beta
e^{i\theta}\left\vert 10\right\rangle =$ $\alpha\left\vert 0_{L}\right\rangle
+\beta e^{i\theta}\left\vert 1_{L}\right\rangle ,$ where the phase $\theta$ is
determined by the characteristic of the potential barrier$.$ Besides its own
function in universal quantum computation, this rotation gate $R_{z}$ will be
also used below for achieving other gates.

Another important single-logic-qubit gate is Hadamard gate, i.e.,
\begin{equation}
\alpha\left\vert 0_{L}\right\rangle +\beta\left\vert 1_{L}\right\rangle
\rightarrow\lbrack\alpha(\left\vert 0_{L}\right\rangle +\left\vert
1_{L}\right\rangle )+\beta(\left\vert 0_{L}\right\rangle -\left\vert
1_{L}\right\rangle )]/\sqrt{2},
\end{equation}
with $\alpha$ and $\beta$ arbitrary numbers, and $\alpha^{2}+\beta^{2}=1.$ To
do this job in DFS, we have to introduce\ ancillary physical-qubits
$1^{\prime}$ and $2^{\prime}$ which are initially prepared in $(\left\vert
01\right\rangle _{1^{\prime}2^{\prime}}+\left\vert 10\right\rangle
_{1^{\prime}2^{\prime}})/\sqrt{2}$. As shown in the inset of Fig. 2, the
initial state $(\left\vert 01\right\rangle _{1^{\prime}2^{\prime}}+\left\vert
10\right\rangle _{1^{\prime}2^{\prime}})/\sqrt{2}$ can be made in the case of
P$_{1^{\prime}2^{\prime}}$=0, where P$_{1^{\prime}2^{\prime}}$ is a charge
parity measurement building block designed in Fig. 2 of \cite{9}.
P$_{1^{\prime}2^{\prime}}$ functions as: P=1 means the two input spins
aligned, and P=0 corresponds to the opposite spins input. So we have the
initial state of the system as%
\begin{align*}
\left\vert \Psi\right\rangle  &  =\left(  \alpha\left\vert 0_{L}\right\rangle
+\beta\left\vert 1_{L}\right\rangle \right)  _{12}\otimes(\left\vert
01\right\rangle _{1^{\prime}2^{\prime}}+\left\vert 10\right\rangle
_{1^{\prime}2^{\prime}})/\sqrt{2}\\
&  =\left[  \alpha\left(  \left\vert 0101\right\rangle _{121^{\prime}%
2^{\prime}}+\left\vert 0110\right\rangle _{121^{\prime}2^{\prime}}\right)
+\beta\left(  \left\vert 1001\right\rangle _{121^{\prime}2^{\prime}%
}+\left\vert 1010\right\rangle _{121^{\prime}2^{\prime}}\right)  \right]
/\sqrt{2}.
\end{align*}
\ Besides the ancillary qubits, a controlled-phase operation between
physical-qubits 1 and $1^{\prime}$ is necessary in constructing our Hadamard
gate for logic-qubits, which is shown in Appendix and also in Fig. 1(b). After
this controlled-phase gate is applied, we have%
\begin{equation}
\left\vert \Psi^{\prime}\right\rangle =\left[  \alpha\left(  \left\vert
0101\right\rangle _{121^{\prime}2^{\prime}}+\left\vert 0110\right\rangle
_{121^{\prime}2^{\prime}}\right)  +\beta\left(  \left\vert 1001\right\rangle
_{121^{\prime}2^{\prime}}-\left\vert 1010\right\rangle _{121^{\prime}%
2^{\prime}}\right)  \right]  /\sqrt{2}.
\end{equation}
Then if we find in Fig. 2 that $D_{1}=1$ and $D_{2}=0$, which implies that
input physical-qubits are in a state $(\left\vert 01\right\rangle +\left\vert
10\right\rangle )/\sqrt{2}$ \cite{9}$,$ then the qubit pair $1^{\prime}%
$-$2^{\prime}$ collapses to
\[
\alpha\left(  \left\vert 01\right\rangle _{1^{\prime}2^{\prime}}+\left\vert
10\right\rangle _{1^{\prime}2^{\prime}}\right)  /\sqrt{2}+\beta\left(
\left\vert 01\right\rangle _{1^{\prime}2^{\prime}}-\left\vert 10\right\rangle
_{1^{\prime}2^{\prime}}\right)  /\sqrt{2},
\]
which is exactly a Hadamard gate for logic-qubits in DFS but with the quantum
information transferred from pair 1-2 to pair 1$^{\prime}$- 2$^{\prime}$. This
trick has also been used previously for neutral atoms in DFS \cite{8}. We may
have an alternative for the last step, that is, in the case of $D_{1}=0$ and
$D_{2}=1,$ the qubit pair $1^{\prime}$-$2^{\prime}$ collapses to%
\[
\alpha\left(  \left\vert 01\right\rangle _{1^{\prime}2^{\prime}}+\left\vert
10\right\rangle _{1^{\prime}2^{\prime}}\right)  /\sqrt{2}-\beta\left(
\left\vert 01\right\rangle _{1^{\prime}2^{\prime}}-\left\vert 10\right\rangle
_{1^{\prime}2^{\prime}}\right)  /\sqrt{2}.
\]
For this situation, we may also have our desired Hadamard gate after
additional spin-flip $\sigma_{x}$ operations are applied on physical-qubits 1'
and 2', respectively. Therefore we have achieved a logic-qubit Hadamard gate
deterministically by four physical-qubits with quantum information transferred
from the pair 1-2 to the pair 1$^{\prime}$-2$^{\prime}$, i.e., $\left\vert
0_{L}\right\rangle _{12}\Longrightarrow(\left\vert 0_{L}\right\rangle
_{1^{\prime}2^{\prime}}+\left\vert 1_{L}\right\rangle _{1^{\prime}2^{\prime}%
})/\sqrt{2}$ and $\left\vert 1_{L}\right\rangle _{12}\Longrightarrow$
$(\left\vert 0_{L}\right\rangle _{1^{\prime}2^{\prime}}-\left\vert
1_{L}\right\rangle _{1^{\prime}2^{\prime}})/\sqrt{2}.$

To accomplish a universal quantum computation, however, we also need a
nontrivial two-logic-qubit gate. Fig. 3 demonstrates a controlled-phase gate
for two logic-qubits consisting of pairs 1-2 and 5-6, assisted by an ancillary
pair 3-4 which are measured at the end of the circuit. So the total system is
initially prepared in%
\begin{equation}
\left\vert \Phi\right\rangle =\left(  \alpha\left\vert 01\right\rangle
_{12}+\beta\left\vert 10\right\rangle _{12}\right)  \otimes(\left\vert
01\right\rangle _{34}+\left\vert 10\right\rangle _{34})/\sqrt{2}\otimes\left(
c\left\vert 01\right\rangle _{56}+d\left\vert 10\right\rangle _{56}\right)  ,
\end{equation}
where $\alpha,$ $\beta,$ c and d are arbitrary numbers with $\alpha^{2}%
+\beta^{2}=1$ and $c^{2}+d^{2}=1.$ After we have the charge detection at
$P_{13}$ with $P_{13}=$1, the system becomes%
\begin{equation}
\left(  \alpha\left\vert 0101\right\rangle _{1234}+\beta\left\vert
1010\right\rangle _{1234}\right)  \otimes\left(  c\left\vert 01\right\rangle
_{56}+d\left\vert 10\right\rangle _{56}\right)  ,
\end{equation}
where the first four qubits are entangled as a Bell state of the logic-qubits
when $\alpha=\beta=1/\sqrt{2}$. Please note that, if $P_{13}=0,$ the system
are still in the DFS, i.e.,%
\[
\left(  \alpha\left\vert 0110\right\rangle _{1234}+\beta\left\vert
1001\right\rangle _{1234}\right)  \otimes\left(  c\left\vert 01\right\rangle
_{56}+d\left\vert 10\right\rangle _{56}\right)  ,
\]
from which we may also achieve Eq. (4) by additional spin-flip operations
$\sigma_{x}$ on physical-qubits 3 and 4. Following the operation steps in Fig.
3, we have to twice use the Hadamard gate designed in Fig. 2, and go through a
block P$_{46}.$ In the case of P$_{46}=1,$ we obtain the output of the system,%
\begin{align}
\left\vert \Phi^{\prime}\right\rangle  &  =\left(  \alpha c\left\vert
0101\right\rangle _{1256}+\alpha d\left\vert 0110\right\rangle _{1256}+\beta
c\left\vert 1001\right\rangle _{1256}-\beta d\left\vert 1010\right\rangle
_{1256}\right)  \left\vert 01\right\rangle _{34}\nonumber\\
&  +\left(  \alpha c\left\vert 0101\right\rangle _{1256}-\alpha d\left\vert
0110\right\rangle _{1256}+\beta c\left\vert 1001\right\rangle _{1256}+\beta
d\left\vert 1010\right\rangle _{1256}\right)  \left\vert 10\right\rangle
_{34}.
\end{align}
So we measure the physical-qubits 3 and 4. The output $\left\vert
01\right\rangle _{34}$ means a two-logic-qubit controlled phase gate to be
achieved. Alternatively, if we have $\left\vert 10\right\rangle _{34}$, the
controlled phase gate is also available after an additional operation
$\sigma_{z}^{5}\otimes I^{6}$ is applied on the pair 5-6. Moreover, if we find
P$_{46}=0,$ the output should be in the state,%
\begin{align}
\left\vert \Phi^{\prime}\right\rangle  &  =\left(  \alpha c\left\vert
0101\right\rangle _{1256}+\alpha d\left\vert 0110\right\rangle _{1256}-\beta
c\left\vert 1001\right\rangle _{1256}+\beta d\left\vert 1010\right\rangle
_{1256}\right)  \left\vert 01\right\rangle _{34}\nonumber\\
&  +\left(  -\alpha c\left\vert 0101\right\rangle _{1256}+\alpha d\left\vert
0110\right\rangle _{1256}+\beta c\left\vert 1001\right\rangle _{1256}+\beta
d\left\vert 1010\right\rangle _{1256}\right)  \left\vert 10\right\rangle
_{34}.
\end{align}
Like previously, if the measurement on physical-qubits 3 and 4 yields
$\left\vert 01\right\rangle _{34}$ ($\left\vert 10\right\rangle _{34}$), we
may achieve the two-logic-qubit controlled phase gate after applying operation
$\sigma_{z}$ on physical-qubits 1 (1 and 5).

\section{Discussion}

It was considered that quantum computation using free electrons is more
powerful than using photons because the former is a deterministic one and thus
needs less ancillary qubits \cite{9}. Moreover, as the charge detection is
non-destructive with respect to the qubits encoded in electron spins, there is
no loss of qubits in the detection. As a result, we could have deterministic
quantum gating by free electrons in the DFS, which is much more efficient than
the probabilistic gating by photons. Furthermore, the ideas with photons - the
flying qubits - interacting with static qubits (such as cold atoms or ions)
are referred to as a promising way towards scalable quantum computation. While
as the information conversion based on the interaction between photons and
matter is generally inefficient, this kind of quantum computation, even in
principle scalable, is of very low success rate. In contrast, the free
electrons, for example, conduction electrons in solid-state materials, could
play roles as both flying and static qubits, which would be a promising
candidate for large-scale quantum computation \cite{16}.

Although the encoding with electron pairs by our scheme makes more than half
numbers of qubits sacrificed, which increases overhead for quantum
computation, the operation carried out in DFS much reduces the
implementational time and difficulty, and enhances the fidelity. As seen in
current experiments, refocusing pulses to remove dephasing have to be applied
repeatedly even for a single quantum gating. In this sense, the sacrifice of
qubit resource in our scheme helps saving time and laser pulses, and thereby
increases the fidelity of implementation. In other words, with the encoding,
we could treat free electrons with the same way as photons for a universal
quantum computation. As our scheme could be done in a deterministic way, which
is the prerequisite for a scalable quantum computation, the quantum
computation with free electrons by our scheme could be more advantageous over
the probabilistic ways by photons.

As mentioned previously, our focus is on the collective dephasing, which is
supposed to be a dominant error due to magnetic field fluctuation and could be
perfectly eliminated by our encoding. While there are also other errors beyond
the collective dephasing one in a real quantum computation, such as logic
errors regarding $\sigma_{r}^{i}\sigma_{r}^{i+1}$ with r=x, y, z, and the
leakage errors related to following undesired operations: $\sigma_{x}^{i}$,
$\sigma_{x}^{i+1}$, $\sigma_{y}^{i}$, $\ \sigma_{y}^{i+1}$, $\sigma_{x}%
^{i}\sigma_{z}^{i+1}$, $\sigma_{z}^{i}\sigma_{x}^{i+1}$, $\sigma_{y}^{i}%
\sigma_{z}^{i+1}$, $\sigma_{z}^{i}\sigma_{y}^{i+1\text{ }}$\cite{14}. To
eliminate the logic errors, we may employ deliberately designed pulse
sequences including $\sigma_{x}^{i}\sigma_{x}^{i+1},$ $\sigma_{y}^{i}%
\sigma_{y}^{i+1},$ and $\sigma_{z}^{i}\sigma_{z}^{i+1}$ respectively. As this
implementation might be not fast enough, however, the 'Bang-Bang' pulse
control would not fully remove the logic errors, which need further amendment
by refocusing on individual physical-qubits \cite{14}. The leakage errors
could be in principle fully removed by the leakage-elimination operator
introduced by \cite{15}. This operator is actually associated with projection
operations which could be applied to our free electron case. Moreover, since
the errors are brought about by unpredictable factors, e.g., the fluctuation
of the magnetic field, we have to first use interrogative 'Bang-Bang' pulses
to determine the required values for correction, which has been actually a
sophisticated technique \cite{17}. A very recent experiment \cite{18} with
two-dimensional electron gas made of GaAs/AlGaAs heterostructure has
successfully extended T$_{2}$ from 10 nanosec to 1 micorsec by using the
'Bang-Bang' pulses. So, with the encoding plus 'Bang-Bang' pulses assisted
sometimes by individual physical-qubit refocusing,\ all dephasing errors could
be strongly suppressed in our free-electron quantum computation \cite{14}.

Besides dephasing, however, there are other sources of decoherence to affect
free electrons, e.g., the decoherence yielding T$_{1}.$ Fortunately, as
reported in \cite{19}, T$_{1}$ could be extended from microsec to tens of
millisec in systems of two-dimensional electron gas by means of 'Bang-Bang'
pulses. This spin relaxing time is long enough for us to achieve high-quality
quantum information processing.

The beam splitters and charge detectors employed in our scheme have been
experimentally achieved by means of the point contacts in a two-dimensional
electron gas \cite{20,21,22,23}. From the experimental values, we may estimate
that each beam splitter takes hundreds of nanosec for an electronic parity
check \cite{20,21} and each Hadamard gate also takes hundreds of nanosec
\cite{24}. Moreover, the currently achievable time resolution for charge
detection is of the order of $micro\sec$ \cite{21}. So considering all the
operations in the gates designed in the present paper, the time we have to
take would be of the order of microsec, which is much shorter than the above
mentioned T$_{1}.$ With the advance of techniques, we believe that this time
would be further reduced in the near future. For example, it is expected that
the time resolved detection required in charge detector for the ballistic
electrons in a semiconductor will be in the $pico\sec$ range \cite{9,25}.
Compared to previous proposals for electron-spin-based quantum computation,
e.g., the famous one \cite{26} in which the qubits are encoded in the spin of
the single access electron in conduction band of semiconductor quantum dots
and the two-qubit gate is carried out by exchange interaction in the interdot
tunneling, our scheme is not more stringent in implementation. Particularly,
as\ T$_{2}$ is much longer in our scheme with DFS encoding, we could achieve a
quantum computation with higher fidelity than by any previous scheme without
the DFS encoding. Furthermore, as our proposal is for free electrons, our idea
could be applied to the spin-based quantum computing proposal for the
electrons floating on liquid Helium \cite{27}.

On the other hand, since the charge detectors in our scheme function to
distinguish the antiparallel states of two electron-spin from\ the parallel
states, we may consider an alternative way, as reported in \cite{10}, to
convert spin parity into charge information by resonant tunneling between two
quantum dots. With this resonant tunneling characteristic, one can determine
the parity of two-qubit spins by differentiating the spin states with
antiparallel spins from the ones with parallel spins, in which quantum
information encoded in the spin qubits is not destroyed.

\section{Conclusion and acknowledgments}

In summary, we have proposed a potential scheme for universal quantum
computation with free electrons in DFS in a deterministic way by using
polarizing beam splitters, charge detectors, and single-spin rotations. Two
noncommutable single-logic-qubit gates, i.e., $R_{z}$ operation and Hadamard
gate, and a two-logic-qubit\ controlled phase gate have been designed and
demonstrated. With these building blocks, an universal dephasing-free quantum
computing architecture can be constructed for free electrons.

This work is supported in part by National Natural Science Foundation
of\ China under Grant Nos. 10474118 and 60490280, by Hubei Provincial Funding
for Distinguished Young Scholars, and by the National Fundamental Research
Program of China under Grant No. 2005CB724502 and No. 2006CB921203.

\[%
\begin{tabular}
[c]{|l|l|l|}\hline
$P_{2}\setminus P_{1}$ & $0$ & $1$\\\hline
$0$ & $R_{z}(\varphi)=e^{i(E_{1}\Delta t+\pi)}$, $R_{z}(\phi)=e^{i(E_{1}\Delta
t^{\prime}+\pi)}$ & $R_{z}(\varphi)=e^{-i(E_{1}\Delta t-\pi)}$, $R_{z}%
(\phi)=e^{iE_{1}\Delta t^{\prime}}$\\\hline
$1$ & $R_{z}(\varphi)=e^{iE_{1}\Delta t}$, $R_{z}(\phi)=e^{-i(E_{1}\Delta
t^{\prime}-\pi)}$ & $R_{z}(\varphi)=e^{-iE_{1}\Delta t}$, $R_{z}%
(\phi)=e^{-iE_{1}\Delta t^{\prime}}$\\\hline
\end{tabular}
\ .
\]

\section{Appendix}

We present detailed derivation of Eq. (2) below, in which the relative phases
due to free evolution of different levels will be fully considered. We set
$t_{0}$ to be the time that physical-qubits 1 and 1$^{\prime}$ go into the
circuit C-R. After a time $\Delta t$ they reach the block $P_{1}$, as shown in
Fig. 1 (b), and the total system turns to,%
\begin{align}
\left\vert \Psi\left(  t_{0}+\Delta t\right)  \right\rangle  &  =\frac
{e^{i2E_{1}\Delta t}}{2}\left[  \alpha\left(  \left\vert 0101\right\rangle
_{121^{\prime}2^{\prime}}+\left\vert 0110\right\rangle _{121^{\prime}%
2^{\prime}}\right)  +\beta\left(  \left\vert 1001\right\rangle _{121^{\prime
}2^{\prime}}+\left\vert 1010\right\rangle _{121^{\prime}2^{\prime}}\right)
\right] \nonumber\\
&  \otimes\left(  \left\vert 0\right\rangle _{a}+\left\vert 1\right\rangle
_{a}e^{iE_{1}\Delta t}\right)  , \tag{(A1)}%
\end{align}
where we have assumed the energies of the states $\left\vert 0\right\rangle $
and $\left\vert 1\right\rangle $ to be, respectively, $0$ and $-E_{1}$. The
subscripts n ($n=1,2,1^{\prime},2^{\prime},a$) represent the corresponding
physical-qubits. Qubits 1 and 1$^{\prime}$ are input from the control-in and
target-in ports of the circuit C-R, respectively. Qubit $a$ is input from the
auxiliary port [See Fig. 1(b)]. Other two qubits, i.e., qubits 2 and
2$^{\prime}$ are idle at this stage. If we have $P_{1}=1,$ the state of the
system collapses into
\begin{align}
&  \left\vert \Psi\left(  t_{0}+\Delta t\right)  \right\rangle =\frac
{e^{i2E_{1}\Delta t}}{\sqrt{2}}[\alpha\left(  \left\vert 0101\right\rangle
_{121^{\prime}2^{\prime}}\left\vert 0\right\rangle _{a}+\left\vert
0110\right\rangle _{121^{\prime}2^{\prime}}\left\vert 0\right\rangle
_{a}\right) \tag{(A2)}\\
&  +\beta\left(  \left\vert 1001\right\rangle _{121^{\prime}2^{\prime}%
}\left\vert 1\right\rangle _{a}e^{iE_{1}\Delta t}+\left\vert 1010\right\rangle
_{121^{\prime}2^{\prime}}\left\vert 1\right\rangle _{a}e^{iE_{1}\Delta
t}\right)  ].\nonumber
\end{align}
Then we perform Hadamard operation on the ancillary qubit during a time length
$\Delta t^{\prime}$, which yields,%
\begin{align}
\left\vert \Psi\left(  t_{0}+\Delta t+\Delta t^{\prime}\right)  \right\rangle
&  =\frac{e^{i2E_{1}(\Delta t+\Delta t^{\prime})}}{2}[\alpha\left\vert
0101\right\rangle _{121^{\prime}2^{\prime}}(\left\vert 0\right\rangle
_{a}+\left\vert 1\right\rangle _{a}e^{iE_{1}\Delta t^{\prime}})\nonumber\\
&  +\alpha\left\vert 0110\right\rangle _{121^{\prime}2^{\prime}}(\left\vert
0\right\rangle _{a}+\left\vert 1\right\rangle _{a}e^{iE_{1}\Delta t^{\prime}%
})\nonumber\\
&  +\beta\left\vert 1001\right\rangle _{121^{\prime}2^{\prime}}(\left\vert
0\right\rangle _{a}e^{iE_{1}\Delta t}-\left\vert 1\right\rangle _{a}%
e^{iE_{1}(\Delta t+\Delta t^{\prime})})\tag{(A3)}\\
&  +\beta\left\vert 1010\right\rangle _{121^{\prime}2^{\prime}}(\left\vert
0\right\rangle _{a}e^{iE_{1}\Delta t}-\left\vert 1\right\rangle _{a}%
e^{iE_{1}(\Delta t+\Delta t^{\prime})})].\nonumber
\end{align}
In the case of $P_{2}$ = 1, we have,
\begin{align}
\left\vert \Psi\left(  t_{0}+\Delta t+\Delta t^{\prime}\right)  \right\rangle
&  =\frac{e^{i2E_{1}(\Delta t+\Delta t^{\prime})}}{2}[\alpha\left(  \left\vert
0101\right\rangle _{121^{\prime}2^{\prime}}\left\vert 0\right\rangle
_{a}+\left\vert 0110\right\rangle _{121^{\prime}2^{\prime}}\left\vert
1\right\rangle _{a}e^{iE_{1}\Delta t^{\prime}}\right) \nonumber\\
&  +\beta(\left\vert 1001\right\rangle _{121^{\prime}2^{\prime}}\left\vert
0\right\rangle _{a}e^{iE_{1}\Delta t}-\left\vert 1010\right\rangle
_{121^{\prime}2^{\prime}}\left\vert 1\right\rangle _{a}e^{iE_{1}(\Delta
t+\Delta t^{\prime})})]. \tag{(A4)}%
\end{align}
Then a measurement on the ancillary qubit in the dressed state \{$\left\vert
+\right\rangle $, $\left\vert -\right\rangle $\} leads to
\begin{equation}
\alpha\left(  \left\vert 0101\right\rangle _{121^{\prime}2^{\prime}%
}+\left\vert 0110\right\rangle _{121^{\prime}2^{\prime}}e^{iE_{1}\Delta
t^{\prime}}\right)  +\beta(\left\vert 1001\right\rangle _{121^{\prime
}2^{\prime}}e^{iE_{1}\Delta t}-\left\vert 1010\right\rangle _{121^{\prime
}2^{\prime}}e^{iE_{1}(\Delta t+\Delta t^{\prime})}), \tag{(A5)}%
\end{equation}
in the case of the output $\left\vert +\right\rangle $. If the output is
$\left\vert -\right\rangle $, an additional operation $\sigma_{z}$ applied to
qubit 1$^{\prime}$ is necessary. Assisted by single-qubit rotations
$R_{z}(\varphi)=e^{-iE_{1}\Delta t}$ and $R_{z}(\phi)=e^{-iE_{1}\Delta
t^{\prime}}$, we could obtain Eq. (2). In fact, as shown in Eq. (A3), there is
a possible leakage out of the DFS in the building blocks of the C-R operation
in Fig. 1 (b). By a selective single-qubit rotations $R_{z}$ in the building
blocks, however, we can suppress this leakage, just like what is done by the
'Bang-Bang' pulse sequences introduced in \cite{14,15}.

For different values of P$_{1}$ and P$_{2},$ the deductions are similar and we
can also reach Eq. (2) after different rotating phases by $R_{z},$ as shown in
Table 1.

\newpage%

%TCIMACRO{\FRAME{ftbpFU}{6.2967in}{9.0615in}{0pt}{\Qcb{{}Fig. 1}}{}%
%{fig1.ps}{\special{ language "Scientific Word";  type "GRAPHIC";
%width 6.2967in;  height 9.0615in;  depth 0pt;  original-width 7.8646in;
%original-height 11.3334in;  cropleft "0";  croptop "1";  cropright "1";
%cropbottom "0";  filename 'Fig1.ps';file-properties "XNPEU";}}}%
%BeginExpansion
\begin{figure}
[ptb]
\begin{center}
\includegraphics[
height=9.0615in,
width=6.2967in
]%
{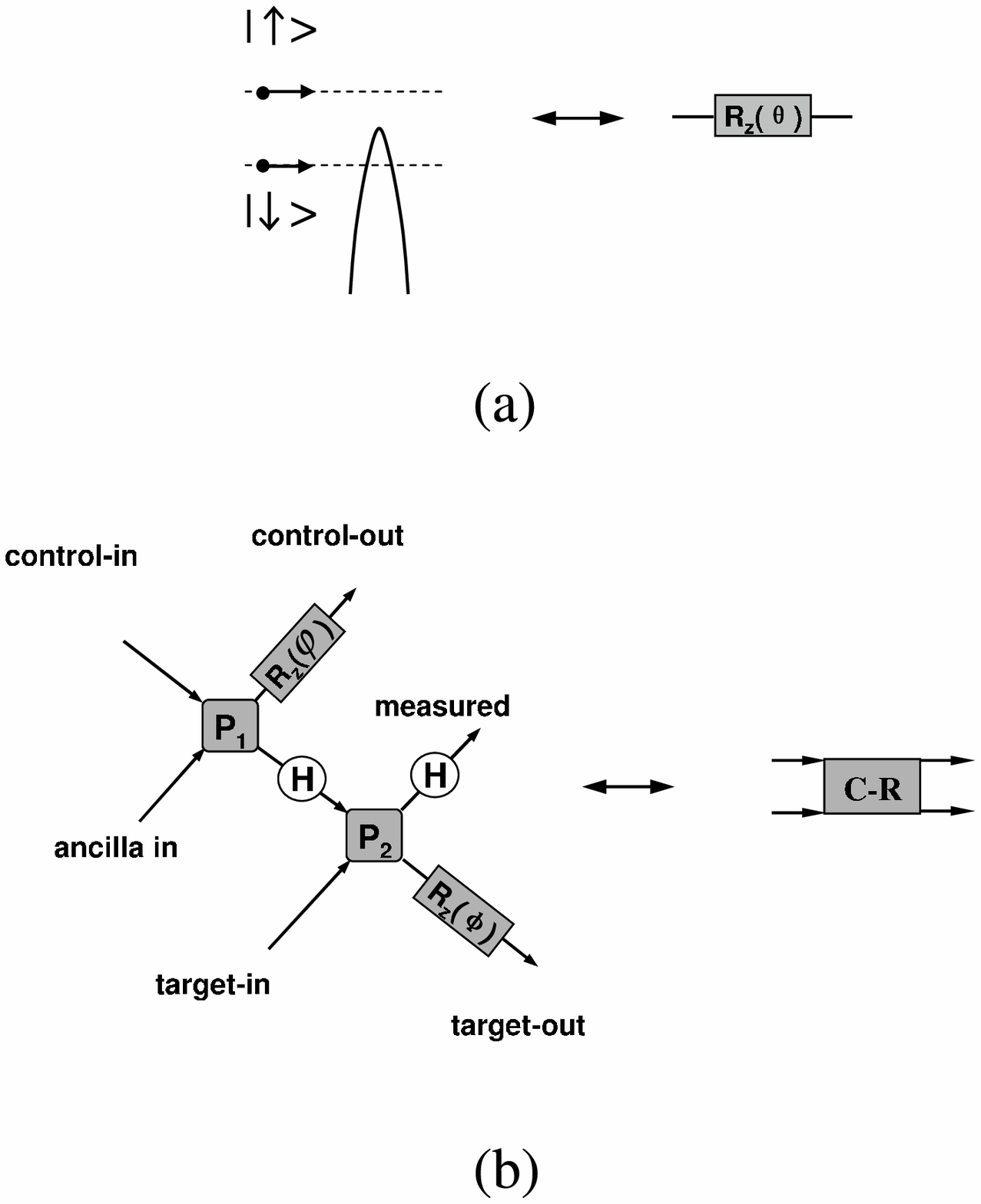}%
\caption{{}Fig. 1}%
\end{center}
\end{figure}
%EndExpansion
\newpage%

%TCIMACRO{\FRAME{ftbpFU}{6.2967in}{9.0615in}{0pt}{\Qcb{{}Fig. 2}}{}%
%{fig2.ps}{\special{ language "Scientific Word";  type "GRAPHIC";
%width 6.2967in;  height 9.0615in;  depth 0pt;  original-width 7.8646in;
%original-height 11.3334in;  cropleft "0";  croptop "1";  cropright "1";
%cropbottom "0";  filename 'Fig2.ps';file-properties "XNPEU";}}}%
%BeginExpansion
\begin{figure}
[ptb]
\begin{center}
\includegraphics[
height=9.0615in,
width=6.2967in
]%
{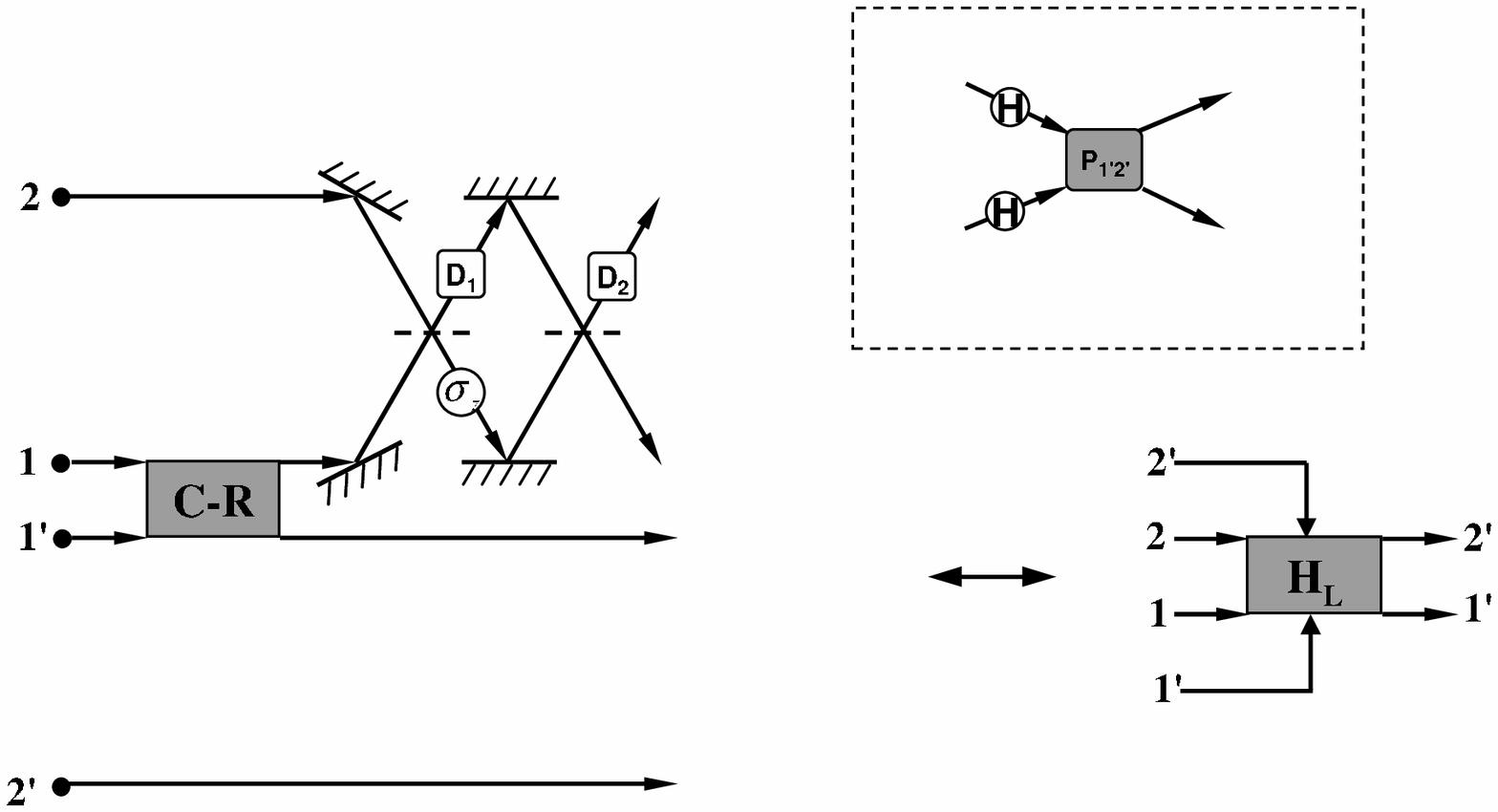}%
\caption{{}Fig. 2}%
\end{center}
\end{figure}
%EndExpansion
\newpage%

%TCIMACRO{\FRAME{ftbpFU}{6.2967in}{9.0615in}{0pt}{\Qcb{{}Fig. 3}}{}%
%{fig3.ps}{\special{ language "Scientific Word";  type "GRAPHIC";
%width 6.2967in;  height 9.0615in;  depth 0pt;  original-width 7.8646in;
%original-height 11.3334in;  cropleft "0";  croptop "1";  cropright "1";
%cropbottom "0";  filename 'Fig3.ps';file-properties "XNPEU";}}}%
%BeginExpansion
\begin{figure}
[ptb]
\begin{center}
\includegraphics[
height=9.0615in,
width=6.2967in
]%
{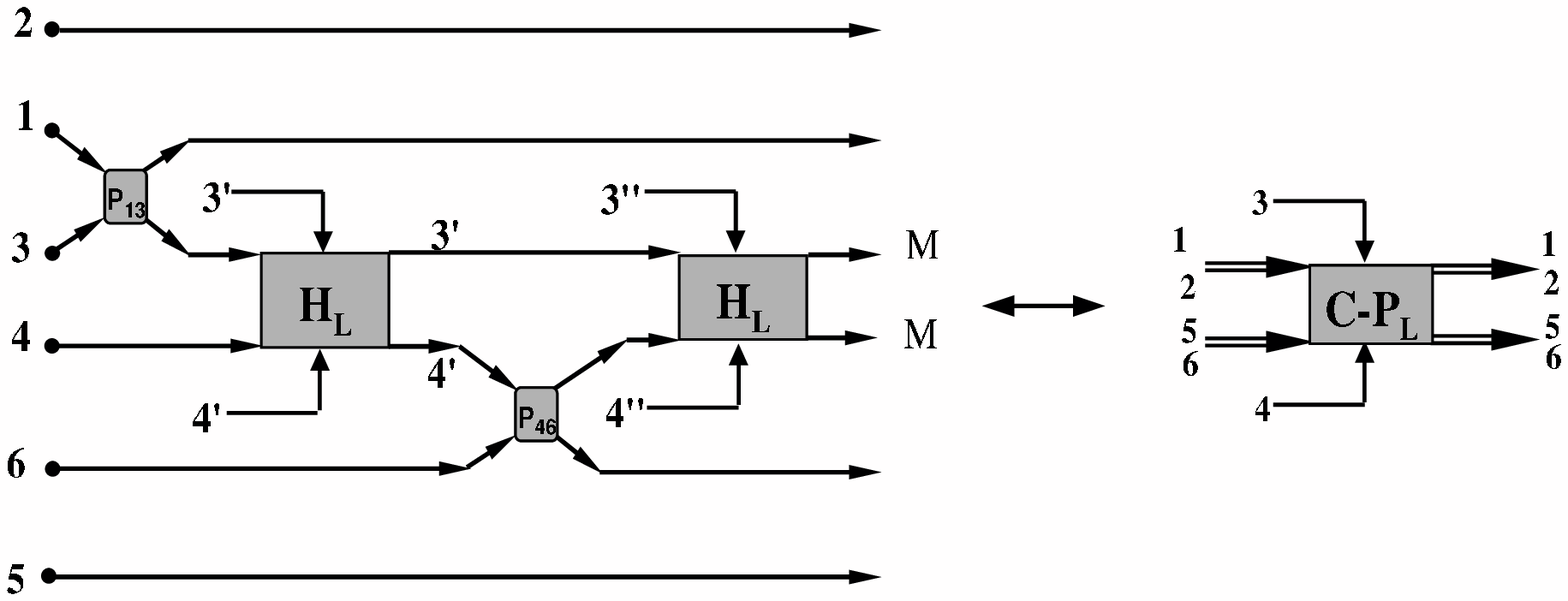}%
\caption{{}Fig. 3}%
\end{center}
\end{figure}
%EndExpansion

\end{document}